\documentclass[aps,superscriptaddress,eqsecnum,nofootinbib,showpacs,preprintnumbers]{revtex4}
\usepackage{graphicx,epsfig}
\usepackage{amsmath}
\usepackage {amssymb}

\newcommand{\be}{\begin{eqnarray}}
\newcommand{\ee}{\end{eqnarray}}
\newcommand{\bea}{\begin{eqnarray}}
\newcommand{\nn}{\nonumber}
\newcommand{\eea}{\end{eqnarray}}

\def\a{\alpha}
\def\b{\beta}
\def\g{\gamma}

\def\d{\delta}

\def\la{\lambda}
\def\La{\Lambda}
\def\k{\kappa}
\def\m{\mu}
\def\n{\nu}
\def\r{\rho}

\def\s{\sigma}

\def\f{\phi}

\def\z{\zeta}

\begin{document}

\title{On Lovelock galileons and black holes}
\author{Christos Charmousis}
\email{christos.charmousis@th.u-psud.fr}
\affiliation{Laboratoire de Physique Th\'eorique (LPT), Univ. Paris-Sud, CNRS UMR 8627, F-91405 Orsay, France}
%\affiliation{Laboratoire de Math\'ematiques et Physique Th\'eorique (LMPT), CNRS UMR 6083, Universit\'e Francois Rabelais-Tours, France}
\author{Minas Tsoukalas}
\email{minasts@cecs.cl} \affiliation{Centro de Estudios
Cient\' ificos, Casilla 1469, Valdivia, Chile}
\preprint{LPT-Orsay-15-38}

\date{\today}

\begin{abstract}

We study a scalar-tensor version of Lovelock theory with a non trivial higher order galileon term involving the coupling of the Lovelock two tensor with derivatives of the scalar galileon field. For a static and spherically symmetric spacetime we extend the Boulware-Deser solution to the presence of a Galileon field. The hairy solution has a regular scalar field on the black hole event horizon and presents certain self tuning properties for the bulk cosmological constant and the Gauss-Bonnet coupling. The combined time and radial dependence of the galileon field permits its horizon regularity. Furthermore in order to investigate the effects of linear time dependence we find spherically symmetric solutions in 4 and 5 spacetime dimensions. They are shown to have singular horizons. Afar from the Schwarzschild radius and for weak higher dimensional couplings the solutions are perturbatively close to GR representing exteriors of GR like star solutions for scalar tensor theories. 

\end{abstract}

\maketitle

%\section{Introduction}

\section{Introduction}
Since a number of years there has been a renewed interest  and  intense activity in the subject of modified gravity. This is true not only at high energy scales but also at infrared scales and very large, cosmological distances. This quite recent intense activity was initiated with the advent of braneworlds where our universe is pictured as a brane in a higher dimensional bulk spacetime. Ultraviolet modifications to 4 dimensional gravity where elegantly predicted in the Randall-Sundrum type of models \cite{RS} with a brane localized 4 dimensional graviton whereas large distance modifications were implemented by the GRS and the DGP models where the 4 dimensional graviton was only quasi-localised becoming of higher dimensional nature at very large distances \cite{GRS}. Although General Relativity with a cosmological constant is the most general metric theory {\footnote{For the precise theorem due to Lovelock \cite{Lovelock:1971yv} and a extensive discussion see for example \cite{charmousis1}}} in 4 dimensions this is not so in higher dimensional theories and implications of this to braneworlds were also investigated (see for example \cite{dufaux}). Implementing 4 dimensional effective theories inspired from braneworld models gave a new perspective to 4 dimensional scalar tensor theories  and also more recently massive gravity and bigravity models \cite{deRham:2014zqa}. In this context while investigating the cosmological constant problem it was noted \cite{fab4} that the most general scalar tensor theory with second order field equations {\footnote{Again for the precise formulation of the theorem one can see for example \cite{Horndeski:1974wa}, \cite{Deffayet:2011gz}. See also interesting generalisations of Horndeski theory in \cite{langlois},\cite{Gao:2014soa},\cite{Padilla:2013jza},\cite{Ohashi:2015fma}.}} was known and written down explicitely by Horndeski \cite{Horndeski:1974wa} back in 1974. In fact, as we will argue now, Horndeski theory is closely related to Lovelock theory \cite{Charmousis:2014mia} and this is the direction we will take in this article. 
  
Lovelock theory is the most general metric theory in arbitrary dimensions with second order field equations \cite{Lovelock:1971yv}. Reducing the theory down to four dimensional spacetimes it gives the most important Horndeski \cite{Horndeski:1974wa}, terms especially all those involving scalar-tensor interactions \cite{VanAcoleyen:2011mj}. Interestingly all four scalar tensor interaction tensor terms are the unique terms of Horndeski theory which self tune in flat spacetime an arbitrary vacuum cosmological constant \cite{fab4}. Furthermore, consistent Kaluza-Klein truncations of Lovelock theory give Galileon black holes \cite{blaise} from their higher dimensional images. The first higher order term of Lovelock theory (after the cosmological constant and the Einstein Hilbert term) is the Gauss-Bonnet term,
\be\label{GB}
{\cal L}_{GB}=\sqrt{-g}\left( R^{2}-4R_{\a\b}R^{\a\b}+R_{\a\b\gamma\delta}R^{\a\b\gamma\delta}\right)
\ee
which is a topological invariant in 4 dimensions. It is also the unique higher order term giving second order field equations in 5 or 6 dimensions. Indeed 
varying (\ref{GB}) with respect to the metric gives us the Lovelock 2-tensor,
\be\label{Hmn1}
H_{\m\n}=-\left[\frac{1}{2}g_{\m\n}\left( R^{2}-4R_{\a\b}R^{\a\b}+R_{\a\b\gamma\delta}R^{\a\b\gamma\delta} \right)-2R\,R_{\m\n}+4R_{\m\rho}R_{\n}^{\,\,\,\rho}+R_{\k\m\rho\n}R^{\k\rho}-2R_{\m\k\la\rho}R_{\n}^{\,\,\,\k\la\rho} \right].
\ee
which is divergence free and identically zero in 4 dimensions giving us in this case the first Lovelock identity (for extensions see for example \cite{Edgar:2001vv,Kastor:2012se,Camanho:2015hea}. 

In four dimensions the action involving the Gauss Bonnet invariant when coupled with a scalar field is no longer a topological invariant. It has a non trivial variation with respect to the scalar field. It is one of the Fab 4 terms, only self tuning with a little help from his friends \cite{fab4}. It is also a very particular term in the Horndeski action as it gives an effective mass term (weighed by spacetime curvature) in the scalar field equation and can bifurcate Galileon no hair arguments \cite{Hui:2012qt}, in the case when the scalar coupling is simply linear for then it still retains translational symmetry \cite{Sotiriou:2013qea}.  Early work adopting perturbative analysis with dilatonic couplings {\footnote{This analysis includes the case of linear coupling}} were performed in \cite{Campbell:1990fu,Campbell:1991rz,Campbell:1991kz,Campbell:1992hc,Kanti:1995cp}, while numerical investigations were performed in \cite{Kanti:1995vq}. Recent investigations on the subject were carried out in \cite{Kaloper:2013vta,Sotiriou:2014pfa}. 
A different way to bifurcate no hair theorems is to involve scalar tensor interactions involving translational invariant Galileons such as the John term of Fab 4 which reads, $G_{\m} \,^{\n} \nabla^{\m}\phi\nabla_{\n}\phi$ where $G_{\mu\nu}$ is the 4 dimensional Einstein tensor. 
There when one considers additionally a linear time dependence of the scalar field \cite{Babichev:2013cya} (see also \cite{Bravo-Gaete:2013dca,Kobayashi:2014eva,Charmousis:2014zaa,Graham:2014ina,Charmousis:2015aya,Babichev:2015rva,Appleby:2015ysa}, and \cite{Kolyvaris:2011fk,Rinaldi:2012vy,Anabalon:2013oea,Minamitsuji:2013ura,Cisterna:2014nua,Cisterna:2015yla,Giribet:2015lfa}) it was shown that the time dependence yielded analytic GR like black holes with additionally well defined scalars on the black hole horizon. 
For a recent review on black holes and scalar fields see \cite{Herdeiro:2015waa} and \cite{Sotiriou:2015pka}.

 In higher dimensions the Gauss Bonnet term as we mentioned earlier is no longer a topological invariant, it is part of the metric theory, allowing for static solutions \cite{Boulware:1985wk,Wheeler:1985nh,Wheeler:1985qd,Cai:2001dz},\cite{charmousis1,Garraffo:2008hu}. As a result, one can construct other than the 4 scalar tensor interactions existing in four dimensions. They have been discussed in the literature but they have not been investigated on the type of solutions or effects they may have \cite{Deffayet:2011gz} (see also \cite{Gaete:2013ixa,Gaete:2013oda,Correa:2013bza,Giribet:2014bva,Giribet:2014fla} ). Indeed the "simplest" of terms can be obtained in a similar way as the John term in Fab 4. It boils down to couple the Lovelock tensor $H_{\m\n}$ with the scalar tensor $\nabla^{\m}\phi\nabla^{\n}\phi$. The divergence freedom of $H_{\m\n}$ guarantees that this is a Galileon term (in other words yields second order equations of motion) and it reads,
\be
\label{johnny}
{\cal L}_{2 HGB}=\sqrt{-g}\, H_{\m} \,^{\n} \nabla^{\m}\phi\nabla_{\n}\phi
\ee
This term has just like Fab 4 terms \cite{fab4}, self-tuning properties although in higher dimensions the application of hiding a bulk cosmological constant is not immediate. Nevertheless it can hide a bulk cosmological constant and we will see this explicitly happening here. 

Our aim in this paper is two fold: first we wish to investigate Lovelock theory including higher order Galileons as (\ref{johnny}) and finding when possible scalar-tensor black hole solutions. We will accomplish this using the construction techniques developed recently \cite{Babichev:2013cya}, \cite{Charmousis:2014zaa},\cite{Babichev:2015rva}. Secondly we want to study in more depth the effect of time dependent linear scalar fields. Towards this aim we will consider the extreme situation where  the scalar field has {\it{only}} time dependence. Here, the scalar or axion is similar to the solutions found recently for three forms in AdS spacetimes in a different physical context more akin to string theory. Can we have regular black holes with a purely time dependent axion? Will there be a black hole horizon as for the axionic solutions of \cite{Bardoux:2012tr}, \cite{Bardoux:2012aw} where the axions are switched in the space-like directions of the horizon? The aforementioned axion solutions  have been discussed in the context of holographic systems as describing momentum dissipation in the boundary of AdS \cite{blaise2}. In fact the free axionic scalars break translational invariance in the horizon coordinates and as a result momentum in the dual three dimensional field theory is no longer conserved. The time dependent axion solutions would then in complete analogy describe energy dissipation for the dual field theory.  We will demonstrate in fact that one has to consider also non trivial radial dependence \cite{Babichev:2013cya} in order to have black hole solutions. We will find that although the solutions have actually a well defined GR limit the scalar tensor solutions are singular, unlike the case of the axionic black holes found previously \cite{Bardoux:2012tr},\cite{Bardoux:2012aw}. We will also show that the solutions are valid GR like star solutions as long as we are far from the Schwarschild radius and for weak higher dimensional coupling.

Keeping these objectives in mind we will consider the following theory,
\be \label{action}
S=\int dx^{d}\sqrt{-g}\big[ \zeta R-2\La+\a \left( R^{2}-4R_{\a\b}R^{\a\b}+R_{\a\b\gamma\delta}R^{\a\b\gamma\delta}\right)-\eta (\partial\phi)^{2}+\b G^{\m\n}\nabla_{\m}\phi\nabla_{\n}\phi+\delta H^{\m\n}\nabla_{\m}\phi\nabla_{\n}\phi\big],
\ee
Notice that the scalar field has translational invariance. So here, we are not in the context of string theory or Kaluza Klein actions where $\phi$ dependence is explicit and the translational invariance is broken. This symmetry will be important for the partial integrability of the system.
We can also see that the dimension of the scalar field is $[\phi]=\frac{d-2}{2}$. For the rest of the coupling constants we have that $[\b]=-2$, $[\d]=-4$ and $\a$ is the usual Gauss-Bonnet coupling with $[\a]=d-4$. This is a rather involved gravitational action but we will see that in order to construct a scalar tensor version of a Boulware-Deser solution all above couplings are necessary. 

Varying the action (\ref{action}) with respect to the metric we have the following equations of motion
\bea\label{metriceqs}
&&\zeta G_{\m\n}-\eta \left(\partial_{\m}\phi\partial_{\n}\phi-\frac{1}{2}g_{\m\n}(\partial \phi)^{2} \right)+g_{\m\n}\La+\a H_{\m\n}\nn\\
&&\qquad\qquad+\beta\left(\frac{1}{2}(\partial \phi)^{2}G_{\m\n}+P_{\m\a\n\b}\nabla^{\a}\phi\nabla^{\b}\phi+\frac{1}{2}g_{\a(\m}\d^{\a\r\s}_{\n)\g\d}\nabla^{\g}\nabla_{\r}\phi\nabla^{\d}\nabla_{\s}\phi \right)\nn\\
&&\qquad\qquad+\delta \Big( -\frac{1}{2}g_{\m\n}H_{\a} \,^{\b} \nabla^{\a}\phi\nabla_{\b}\phi+H_{(\m} \,^{\r} \nabla_{\n)}\phi\nabla_{\r}\phi-\frac{1}{2^{2}}\d^{\lambda \a_{1}\a_{2}\a_{3}\a_{4}}_{\k \b_{1}\b_{2}\b_{3} (\m}R^{\b_{3}}_{\,\,\,\,\,\n)\a_{3}\a_{4}}R^{\b_{1}\b_{2}}_{\a_{1}\a_{2}}\nabla^{\k}\phi\nabla_{\lambda}\phi\nonumber \\
&&\qquad \quad \quad-\frac{1}{2^{2}}\delta^{\lambda \a_{1}\a_{2}\a_{3}\a_{4}}_{\k \b_{1}\b_{2}\b_{4} (\m}g_{\n)\a_{4}}R^{\b_{1}\b_{2}}_{\a_{1}\a_{2}}R_{\a_{3}}\,^{\r\b_{4}\k}\nabla_{\r}\phi\nabla_{\lambda}\phi-\frac{1}{2}\delta^{\lambda \a_{1}\a_{2}\a_{3}\a_{4}}_{\k \b_{1}\b_{2}\b_{4} (\m}g_{\n)\a_{4}}R^{\b_{1}\b_{2}}_{\a_{1}\a_{2}}\nabla^{\k}\nabla_{\a_{3}}\phi\nabla^{\b_{4}}\nabla_{\lambda}\phi\Big)=0,
\eea
where $P_{\a\b\m\n}$ is the double dual of the Riemann tensor, $P_{\a\b\m\n}=-\frac{1}{4}\epsilon_{\a\b\r\s}R^{\r\s\g\d}\epsilon_{\m\n\g\d}$, which can equivalently be written as $P^{\a\b}\,_{\m\n}=\frac{1}{4}\d^{\a\b\la\s}_{\m\n\g\d}R_{\la\s}^{\g\d}$. Expanding the previous expression we have that
\be
P_{\a\b\m\n}=R_{\a\b\m\n}+R_{\b\m}g_{\a\n}-R_{\b\n}g_{\a\m}-R_{\a\m}g_{\b\n}+R_{\a\n}g_{\b\m}+\frac{1}{2}Rg_{\a\m}g_{\b\n}-\frac{1}{2}Rg_{\b\m}g_{\a\n}
\ee
The variation of the action with respect to $\phi$ can be rewritten in the form of a conserved current and this is a consequence of the shift symmetry of
the action,
\be\label{scaleq}
\nabla_{\m}J^{\m}=0, \qquad J^{\m}=\left(\eta g^{\m\n}-\b G^{\m\n}-\d H^{\m\n} \right)\partial_{\n}\phi.
\ee
Note in the above current the presence of the Einstein-Gauss-Bonnet plus cosmological constant metric equations of motion. This is a critical condition in order to find analytic solutions of the scalar tensor system. 

In the next section we will consider purely time dependent scalar fields in 4 and 5 spacetime dimensions. In 4 dimensions the above action boils down to a rather simpler one since the Gauss Bonnet and the Lovelock tensor do not yield any part in the field equations, i.e. $\alpha=\delta=0$. After finding several solutions we will argue that these can never be black hole solutions unlike the case of magnetic axions \cite{Bardoux:2012aw}. We will then go ahead to 5 dimensional scalar tensor Lovelock theory and there find explicit black holes with a non trivial and time dependent scalar generalizing the celebrated Boulware Deser solution \cite{Boulware:1985wk}. We will summarize our results in the last section.

\section{A time dependent axion and the generic irregularity of the solutions}

We want to examine the existence of spherically symmetric static metric solutions where there is only a linear time-dependence in the scalar field. This is a case akin to the electric axionic black holes found recently in \cite{Bardoux:2012tr},\cite{Bardoux:2012aw} or to cosmological galileons where linear dependance is customary (see for example \cite{gilles}. 
The Galileon in this case is an axion switched on in the Killing time direction. It is easy to see that in 4 dimensions its dual three form is in fact magnetic. Starting from 4 dimensions the action (\ref{action}) reduces to,  
\be \label{4daction}
S=\int dx^{4}\sqrt{-g}\big[ \zeta R-2\La-\eta (\partial\phi)^{2}+\b G^{\m\n}\nabla_{\m}\phi\nabla_{\n}\phi\big].
\ee

In \cite{Babichev:2013cya} a linear time-dependent and radial scalar field was considered and regular black hole solutions were found. Here we consider,
\bea
ds^{2}&=&-h(r)dt^{2}+\frac{1}{f(r)}dr^{2}+r^{2}d\Omega_{2}^{2}\\
\phi(t,r)&=&m\, t.
\eea
Doing so the scalar field equation is automatically satisfied. Additionally there is no off-diagonal term in the metric field equations. Hence we are only left with the $(rr)$ and $(tt)$ equations for our two remaining variables to be determined, $f$ and $h$. The $(rr)$ equation reads,
\be\label{4drreq}
\frac{r f h'}{h}=-r^2 \frac{2h \Lambda-m^2 \eta}{2 \zeta h-\beta m^2}-(f-1)\frac{2 \zeta h+\beta m^2}{2 \zeta h-\beta m^2}
\ee 
whereas the $(tt)$ equation reads,
\be\label{4dtteq}
[r(f-1)]'=-r^2 \frac{2h \Lambda+m^2 \eta}{2 \zeta h-\beta m^2}
\ee

In the following we will examine various cases for the values of the couplings, always supposing that the coupling constant $\b$ is non-vanishing. The system of equations is not integrable but we can make some qualitative remarks. We remind the reader that the $\zeta$ coefficient denotes the "GR" - part of the field equations. Then as we can see the effect of the Galileon term is to "shift" the zero's of $h$. This as we will see leads to singular spacetimes as possible zero's of $f$ are not zero`s of $h$.

\subsection{Four dimensional solutions}

Let us start by 
\be
\eta=\Lambda=0\,\,\text{and}\,\,\z\neq0.
\ee 
which is integrable.
In this case from the $(tt)$ equation we get, 
\be
\left(2\z h-m^{2}\b \right)(-1+f+r f')=0.
\ee
If $h(r)=\frac{m^{2}\b}{2\z}$ then we need to have $f(r)=1$. When the second bracket is zero we have
\be
(-1+f+r\,f')=0\,\,\Rightarrow\,\, f(r)=1-\frac{\mu}{r}.
\ee
On the other hand the $(rr)$ equation can be written as,
\be
\int d h \; \frac{2 \zeta h-\beta m^2}{(2 \zeta h+\beta m^2)h}=\int df \frac{1}{f}
\ee
which can be easily integrated and solved for h. We thus have,
\be
(2\z h+m^{2}\b)^{2}=cfh,
\ee
where $c$ is an integration constant that fixes the asymptotics of the solution. Clearly as $f=1$ for large $r$ we need to fix $c=(2\zeta+m^2 \beta)^2$ so that $h=1$ for large $r$. The solution for $h$ reads (only the upper branch is valid),
\be
h=\frac{cf-4m^{2}\b\z+\sqrt{c f}\sqrt{cf-8m^{2}\b\z}}{8\z^{2}}.
\ee 
for the above value of $c$.
Here we note that taking $\beta m^2\rightarrow 0$ the solution smoothly flows to the GR Schwarzschild solution. Hence we are in the same branch as for GR. This is unlike the stealth black hole found in \cite{Babichev:2013cya} where the branch is distinct (although the metric is identical to GR!). However although the solution here is connected to the GR solution it is not a black hole for when we hit the event horizon at $f=0$ we have $h=-\frac{m^2 \beta}{2 \zeta}$. The solution is therefore singular since It will always develop a branch singularity at the zero of the square root when $f<0$. But as long as we are far from the Schwarzschild radius, as is the case for a spherically symmetric star and $\frac{m^2 \beta}{2 \zeta}<<1$ the solution will look like an exterior GR star solution.

The generic case where all coupling constants are not zero is not integrable. However note from the $(tt)$ equation that setting $\frac{\z\,\eta}{\b}+\La=0$ yields a common root on the RHS giving immediately,
\be
f(r)=1-\frac{\mu}{r}+\frac{\eta}{3\b}r^{2}.
\ee
Now we can proceed as in the previous case since the $(rr)$ equation reads,
\be
\int d h \; \frac{2 \zeta h-\beta m^2}{(2 \zeta h+\beta m^2)h}=\int \frac{dr}{rf} (1-f+\frac{\eta}{\b}r^{2})
\ee
and can be written again as
\be
\int d h \; \frac{2 \zeta h-\beta m^2}{(2 \zeta h+\beta m^2)h}=\int df \frac{1}{f}.
\ee
The solution for $h(r)$ comes from solving the algebraic equation
\be
\frac{(2\z h+m^{2}\b)^{2}}{h}=cf,
\ee
where $c$ is an integration constant, or
\be
h=\frac{cf-4m^{2}\b\z\pm\sqrt{c f}\sqrt{cf-8m^{2}\b\z}}{8\z^{2}}.
\ee 
Again we conclude as before that the solutions are not black holes, although again they have a smooth GR limit. 

Since we are not able to analytically  solve the field equations for the full set of coupling constants it is instructive to adopt a perturbative approach in order to find the leading order behavior of the solutions. This is going to be performed by expanding the metric functions in terms of the dimensionless parameter $\epsilon=\frac{m^{2}\b}{2\z}$, supposing that $\epsilon<<1$ and solving order by order the field equations. Additionally we will suppose for simplicity that $\eta=0$. We will thus have
\be
f(r)=f_{0}(r)+\epsilon f_{1}(r)+O(\epsilon^2) \quad\text{and}\quad h(r)=h_{0}(r)+\epsilon h_{1}(r)+ O(\epsilon^2)\quad\text{where}\quad \epsilon=\frac{m^{2}\b}{2\z}<<1
\ee
From the zeroth order term we find that $f_{0}(r)=h_{0}(r)=1-\frac{\m}{r}-\frac{r^{2}\La}{3\z}$, resulting in an (A)dS spacetime, as expected. Going to first order we find that
\bea
&&(r f_{1}(r))'=-\frac{r^{2}\La}{\z f_{0}(r)}\label{f1}\\
&&\left(\frac{h_{1}(r)}{f_{0}(r)}\right)'=\left(\frac{f_{1}(r)}{f_{0}(r)}\right)'-\frac{2\left(f_{0}(r)-1\right)}{r f_{0}(r)^{2}}.
\eea
Integrating the right hand side of (\ref{f1}) we find
\be
\int\frac{3 r^{3}\La}{-3r\z+r^{3}\La+3\z\m}=3 r+9\z\sum_{x_{i}}\frac{(x_{i}-m)\ln(r-x_{i})}{3\La x_{i}^{2}-3\z},
\ee
where $x_{i}$ are the roots of the polynomial
\be
\La x^{3}-3\z x+3\z \m=0.
\ee
This is sufficient to see that our approximation breaks down when we approach any zero of $f_0$ since $f_1$ becomes clearly dominant over $f_0$ close to the horizon. Therefore we come to the same conclusion as we did for our specific analytic examples. The solutions have singular horizons. 

\subsection{Five dimensional case}

For completeness we will examine the case of purely time-dependent scalar field for the full action (\ref{action}). We start by choosing a five dimensional static metric of the form
\be
ds^{2}=-h(r)dt^{2}+\frac{1}{f(r)}dr^{2}+r^{2}d\Omega^{2}_{(3)_{\k}},
\ee
where $d\Omega^{2}_{(3)_{k}}$ represents a three-dimensional maximally symmetric space, where $\k = -1,0,1$ is the spatial curvature of the three-dimensional hypersurfaces, while for the scalar field we have
\be
\phi(t,r)=m\, t.
\ee
The $(tt)$ equation reads,
\be
-6(m^{2}\d-2\a h)[(2\k-f)f]'+2r\big(3\k (m^{2}\b-2\z h)+r^{2}(m^{2}\eta+2\La h) \big)-3(m^{2}\b-2\z h)[r^{2}f]'=0
\ee
and the $(rr)$ equation reads,
\be
-12(\k-f) f (m^{2}\d-2\a h)h'+2rh\big(-3(\k-f)(m^{2}\b+2\z h)+r^{2}(2\La h-m^{2}\eta) \big)-3r^{2}f(m^{2}\b-2\z h) h'=0.
\ee
It is easy to see that setting $\a=\d=0$ we end up in the higher dimensional version of (\ref{4drreq}) and (\ref{4dtteq}). We now have the following constants $\La,\z,\a,\eta,\b,\d$ and $\k$ for the spatial curvature. Setting all coupling constants to zero apart from $\d$ leads to trivial solutions for $f(r)$ and $h(r)$, (mind also that if  $f(r)=\k$, then $h(r)$ is undetermined). Let's examine the following case.

We set:
\be
\z\eta+\b\La=0\,\, \text{and}\,\, \a\eta+\d\La=0.
\ee
The $(tt)$ equation gives,
\be
(m^{2}\eta+2\La h)(2 r^{3}\eta+6r\b\k-6r\b f-3(r^{2}\b+4\d\k-4\d f)f')=0
\ee
while the $(rr)$ equation becomes
\be
2 r(r^{2}\eta+3\b\k-3\b f) (m^{2}\eta-2\La h)\,h+3 f(r^{2}\b+4\d\k-4\d f)(m^{2}\eta+2\La h)h'=0.
\ee
The solution for $f(r)$ is,
\be
f(r)=\k+\frac{r^{2}\b}{4\d}\pm\sqrt{\frac{r^{4}(3\b^{2}-4\d\eta)}{48\d^{2}}+f_{0}}.
\ee
whereas as before $h$ is solved by the quadratic,
\be
(h-\frac{m^2 \eta}{2\Lambda})^2=c f h
\ee
and as before we can conclude that there can never be a horizon covering any of the branch or central singularity at $r=0$. It is absolutely essential that the scalar field has to have a radial profile. This not only regularizes the geometry but also the horizon behavior of the scalar. We move onto this now for the case of Lovelock galileons in $D=5$ dimensions.

\section{Scalar tensor Lovelock black hole}

After examining the axion time dependence on static scacetimes we will now proceed to the case where the scalar field has additionally radial dependence. As we will see the combined time and space dependence of the scalar will regularize it on the horizon. We will look for black holes to the full Lovelock Galileon action (\ref{action}) in 5 dimensional spacetime.

We choose the same static metric Anzatz as before
\be
ds^{2}=-h(r)dt^{2}+\frac{1}{f(r)}dr^{2}+r^{2}d\Omega^{2}_{(3)_{\k}},
\ee
We suppose that the scalar field reads, $\phi=\f(t,r)$ ie., it does not have the same symmetries as the spacetime metric. For the above ansatz the only non-vanishing current components of $J^{\m}$ are the following:
\be
 J^{t}&=&\frac{-\dot{\f}}{2r^{3}h}\left(2r(r^{2}\eta+3\b\k-3\b f)-3(r^{2}\b+4\d\k-4\d f)f' \right)\\
 J^{r}&=&\frac{f\,\f'}{2r^{3}h}\left(2r(r^{2}\eta+3\b\k-3\b f)h-3f(r^{2}\b+4\d\k-4\d f)h' \right),\label{Jr}
\ee
where $( \, \dot{} \, )$ and $(\, '\,)$ represent differentiation with respect to $t$ and $r$ respectively. We see that, due to translation invariance,  the scalar field equation can be written as
\be
\nabla_{\m}J^{\m}=\frac{\partial J^{t}}{\partial t}+\frac{\partial J^{r}}{\partial r}+\frac{1}{2}J^{r}\left(\frac{6}{r}-\frac{f'}{f}+\frac{h'}{h} \right)=0.
\ee
For a generic scalar we see that the scalar field equation can not be written as a first integral and we also end up with a PDE system, which even in the simpler case of \cite{Babichev:2013cya}, is not integrable \cite{Appleby:2015ysa}. Only when the scalar has a linear time dependence, we have a system of ODE s with no explicit time derivatives (but in the presence of the velocity charge $q$). 
In fact one can show that this is the only case where we have consistency and integrability of the system of ODE's \cite{Babichev:2015rva}. The energy momentum tensor associated to the scalar field then does admit a time like Killing vector as it should. 
We therefor have,
\be\label{phi ansatz}
\phi(t,r)=q t+\psi(r).
\ee
Due to the linear time-dependence of the scalar field, there is still an off-diagonal part $(tr)$ in the metric field equations which is not however time dependent. As was proven in \cite{Babichev:2015rva} for (\ref{phi ansatz}) setting this off-diagonal part  to zero is equivalent to the metric satisfying  the following equation
\be\label{EGBrr}
\eta g^{rr}-\b G^{rr}-\d H^{rr}=0.
\ee
The above, (\ref{EGBrr}), immediately satisfies the scalar field equation (\ref{scaleq})  as $J^{r}=0$, \cite{Babichev:2013cya},\cite{Charmousis:2014zaa},\cite{Babichev:2015rva}. Therefore (\ref{phi ansatz}) solves 2 of the field equations at the same time rendering the whole system mathematically consistent. This means that (\ref{EGBrr}) can be solved for $f(r)$ finding that,
\be \label{f expr}
f(r)=\frac{6 r \b h+3\left(r^{2}\b+4\d\k \right)h' \pm \sqrt{-96r\d\left(3\b\k+r^{2}\eta \right)h\,h'+9\left(2 r \b h+\left( r^{2}\b+4\d\k\right)h' \right)^{2}}}{24\d h'},
\ee

We can now use this expression in order to find a relation between the scalar field function $\psi$ and the function $h$. Indeed if we substitute (\ref{phi ansatz}) and (\ref{f expr}) to the $(rr)$ component of (\ref{metriceqs}) we can solve for $\psi'$. The expression is highly non trivial involving a higher order algebraic equation for $\psi'$. But it can still be done and in principle we can replace $\psi'$ in the $(tt)$ equation obtaining a non linear ODE for $h$ which can be eventually solved numerically in general.

Here to go further analytically we choose to set $f(r)=h(r)$ in the metric and search for solutions in the manner of the self-tuning de Sitter Schwarszchild solutions found in \cite{Babichev:2013cya}. From (\ref{EGBrr}), it is easy to see that the solution reads,
  \be \label{BDsol}
f(r)=h(r)=\k+\frac{r^{2}\,\b}{4\,\d}\pm\frac{1}{4}\sqrt{\frac{r^{4}\,(3\b^{2}-4\d\,\eta)}{3\,\d^{2}}+16c_{1}}.
\ee 
Now from the  $(rr)$ component of (\ref{metriceqs}) we can solve for $\psi'$, substitute it back to the equations (alongside with $\psi''$) and see that  the field equations are satisfied, if and only if the coupling constants  satisfy the following relations:
\bea
\d(q^{2}\,\b+\z\k)-\a\,\b\,\k&=&0\label{con1}\\
\eta (\z\k-q^{2}\b)+\La \b \k&=&0\label{con2}
\eea
as well as,
\be
\psi'(r)=\pm\frac{q\sqrt{(\k-f)/\k}}{f},
\ee 
where we have used (\ref{EGBrr}) and  (\ref{con1}),(\ref{con2}) in order to solve for $\a$ and $\La$. It is obvious that eliminating $\k$ from (\ref{con1}) and (\ref{con2}) we get $\eta(\a\b-2\d\z)-\b\d\La=0$.

This solution is very similar to the Boulware Deser metric solution for Einstein Gauss-Bonnet theory with a cosmological constant \cite{Boulware:1985wk,Wheeler:1985nh,Wheeler:1985qd,Cai:2001dz}. Indeed setting $\alpha_{eff}=\delta/\beta$ as an effective Gauss-Bonnet coupling and $\Lambda_{eff}=-6k_{eff}^2=-\eta/\beta$ we obtain,
   \be \label{BDsol1}
f(r)=h(r)=\k+\frac{r^{2}}{4\,\a_{eff}}\left(1\pm\sqrt{1-8\a_{eff}k_{eff}^2+\frac{16\a_{eff} \mu}{r^4}}\right).
\ee 
where $\mu$ is the integration constant associated to mass. We see therefore that the effective higher dimensional coupling and the effective cosmological constant are rescaled via the definitions (\ref{con1}-\ref{con2}) by the scalar velocity parameter $q$. The above solution for the scalar field is valid for non planar horizons. For the particular case where $\k=0$, in order to have a solution we need to have that  $\eta=\frac{3\b^{2}}{4\d}$ and $\a \b\eta =2\d\z\eta+\b\d\La$ and then the scalar field has the form $\psi'(r)=\pm\sqrt{q^{2}+\frac{\d\z-\a\b}{\b\d\,f(r)}}$.

It is obvious that on the horizon of the black hole the radial component $\psi$ blows up but not the scalar field $\phi$. Indeed in ingoing Eddington-Finkelstein coordinates
\be
u=t+\int \frac{1}{f(r)}dr
\ee
the scalar field reads,
\be
\psi'(r)=\pm\frac{q}{\k\pm \k\sqrt{(1-\frac{f}{\k})}}.
\ee
We can now see that when $f(r)$ vanishes the scalar field is regular in the future black hole event horizon (for the plus sign). A different sign choice renders the solution regular at the white hole horizon. %we find

If we now demand that the scalar field has no time dependence, in other words taking the limit $q\rightarrow0$, we see that $\phi(r)=const$. In order to examine the full spectrum of solutions for the scalar static case we must go back to the original equations and solve them again from scratch. 

We will again examine the case where $f(r)=h(r)$. The only non vanishing component of $J^{\m}$ is (\ref{Jr}), and the scalar field equation admits a first integral. 
 Following \cite{Rinaldi:2012vy,Anabalon:2013oea} we can set the integration constant to zero and thus have a single equation which we solve for $f(r)$, giving us again as a solution (\ref{BDsol}). We are now left with the metric field equations.
We find two solutions,
\bea
\a&=&\frac{2\d\z\eta+\b\d\La}{\b\eta}\quad\text{and}\quad\k=0 \quad\text{and thus} \quad\phi'(r)=\sqrt{\frac{\d\z-\a\b}{\b\d\,f(r)}}\\
%&\text{or}&\nn\\
%\a&=&\frac{\d\z}{\b}\quad\text{and}\quad\z\eta+\b\La=0\quad\text{and thus} \quad\phi'(r)=0\\
&\text{or}&\nn\\
\a&=&\frac{2\d}{3\b^{2}}(3\b\z+2\d\La)\quad\text{and}\quad\eta=\frac{3\b^{2}}{4\d}\quad\text{and thus} \quad\phi'(r)=\sqrt{\frac{\d\z-\a\b}{\b\d\,f(r)}}
\eea
As we see in both cases the solution for the scalar field is the same whereas the metric satisfies (\ref{BDsol}) for the relevant constants. Combining the constraints of the second case we see that we can derive the constraint of the first solution. Still that case does not in any way imply the relation $\eta=\frac{3\b^{2}}{4\d}$ and it is only valid for flat horizons, whereas the second solution is valid for any $\k$. Still  the relation $\eta=\frac{3\b^{2}}{4\d}$  for non zero $q$ is mandatory in order to have solutions for flat horizons. Additionally we see that  when $\d\z-\a\b=0$ the scalar field is constant and this is precisely the case mentioned already where $q\rightarrow0$, since then the constraints (\ref{con1}),(\ref{con2}) produce the aforementioned relation.
It is clear that when we abandon time-dependence, the scalar field blows up on the horizon of the black hole as was expected,\cite{Babichev:2013cya}.

\section{Conclusions}

In this paper we have found and studied spherically symmetric solutions of higher order scalar tensor theories. We have investigated not only  4 dimensional Horndeski theories but also theories including higher dimensional galileons stemming from derivative scalar couplings with the Lovelock tensor (\ref{johnny}). We dubbed these Lovelock-galileon terms. With the help of these we have found a regular scalar tensor extension of the Boulware Deser solution \cite{Boulware:1985wk,Wheeler:1985nh,Wheeler:1985qd,Cai:2001dz}. The solution has the basic horizon structure of the afore mentioned solution but with a time and radially dependent scalar field which is furthermore regular at the black hole event horizon. As noted before,  \cite{Babichev:2013cya},%\cite{Charmousis:2014zaa},\cite{Charmousis:2015aya},\cite{Babichev:2015rva},
 this regularity feature is due to the linear time dependence of the scalar field. 
To investigate the role of time dependence of the scalar field further, we also studied a scalar field with only time dependence. In other words although our spacetime has a Killing time like vector the scalar field does not have this symmetry. We found spherically symmetric star solutions but they always have a singular event horizon, practically the scalar shifts the zero of one the metric functions but not the other thus producing a singular space time. Far from the event horizon radius solutions are GR like and asymptotically flat. Furthermore, as we let the scalar velocity charge go to zero we recover GR solutions with or without a cosmological constant. 

The scalar tensor version of the Boulware-Deser solution presents certain self tuning features reminiscent to those of de Sitter black hole solutions \cite{Babichev:2013cya}, \cite{Charmousis:2015aya}. Indeed we see that the bulk cosmological constant as well as the Gauss-Bonnet coupling are rescaled by the scalar velocity charge $q$. This could have interesting consequences in a Randall-Sundrum type braneworld scenario where the fine tuning of the brane tension with the bulk cosmological constant could be relaxed providing a 5 dimensional and regular self tuning solution. But in order to investigate such scenarios we would need to carefully study junction conditions and in particularly in the presence of higher order scalars terms.

~\\

{\bf Acknowledgments}:~~We would like to thank Mokhtar Hassaine, for early collaboration on the subject and for useful discussions. We would also like to thank Elias Kiritsis and Vishagan Sivanesan for discussions. We are also grateful to Georgios Pastras for using his computer for various calculations. The authors thank the National Technical University of Athens for hospitality during the last stages of this work and MT thanks the Laboratoire de Physique Th\'eorique (LPT), Univ.Paris-Sud in Orsay whereas CC thanks The Centro de Estudios Cientificos for hospitality in the initial stages of this work. The authors also acknowledge financial support from the research program, «  Programme national de cosmologie et Galaxies» of the CNRS/INSU, France. The Centro de Estudios Cientificos (CECs) is funded by the Chilean Government through the Centers of Excellence Base Financing Program of Conicyt.

\begin{appendix}

\section{Generalized deltas}

Here we give some definitions on generalized deltas. For more info look in \cite{LovRund} 

\be
\d^{\a_{1}\a_{2}\dots\a_{n}}_{\b_{1}\b_{2}\dots\b_{n}}=n!\d^{\a_{1}}_{[\b_{1}}\d^{\a_{2}}_{\b_{2}}\dots\d^{\a_{n-1}}_{\b_{n-1}}\d^{\a_{n}}_{\b_{n}]},
\ee
where in denoting the antisymmetry we include the normalizing factor, for example
\be
A_{[\m\n]}=\frac{1}{2}(A_{\m\n}-A_{\n\m})
\ee
 This way for example
 \be
\d^{\a_{1}\a_{2}}_{\b_{1}\b_{2}}=2!\d^{\a_{1}}_{[\b_{1}}\d^{\a_{2}}_{\b_{2}]}=\d^{\a_{1}}_{\b_{1}}\d^{\a_{2}}_{\b_{2}}-\d^{\a_{1}}_{\b_{2}}\d^{\a_{2}}_{\b_{1}}.
 \ee
Now making contractions we can define generalized deltas with fewer indices, but the dimensionality of the space that we are working on, plays a crucial role on the value of the numerical factors that appear, after the contraction has been made. So following \cite{LovRund} we have that
\be
\d^{\a_{1}\a_{2}\dots\a_{s}\a_{s+1}\dots\a_{n}}_{\b_{1}\b_{2}\dots\b_{s}\a_{s+1}\dots\a_{n}}=\frac{(d-s)!}{(d-n)!}\d^{\a_{1}\a_{2}\dots\a_{s}}_{\b_{1}\b_{2}\dots\b_{s}}.
\ee
Contracting fully the indices gives us
\be
\d^{\a_{1}\dots\a_{n}}_{\a_{1}\dots\a_{n}}=\frac{d!}{(d-n)!}.
\ee

\section{Lovelock Derivative Couplings}

It is easy to extend the derivative couplings for any Lovelock invariant. These type of couplings belong to the general type of higher order derivative couplings which can give second order field equations and extend Hordenski's theory in arbitrary dimensions \cite{Deffayet:2011gz}. Each one of the Lovelock invariants can be written as
\be
\mathcal{L}_{2n}=\sqrt{-g}\frac{1}{2^{2n}}\delta^{\a_{1}\a_{2}\cdots \a_{2n-1}\a_{2n}}_{\b_{1}\b_{2}\cdots \b_{2n-1}\b_{2n}}\,R_{\a_{1}\a_{2}}^{\b_{1}\b_{2}} \cdots R_{\a_{2n-1}\a_{2n}}^{\b_{2n-1}\b_{2n}}
\ee
If we consider the above term in $d=2n$ then this is a topological invariant and is not dynamical. For example for $n=2$ we have (\ref{GB}), which is a topological invariant in four dimensions. Varying with respect to the metric gives us the following expression
\be\label{Lovetensor}
H_{\m}^{\,\,\,\n}=-\frac{1}{2^{n+1}}\delta^{\n\a_{1}\a_{2}\cdots \a_{2n-1}\a_{2n}}_{\m\b_{1}\b_{2}\cdots \b_{2n-1}\b_{2n}}\,R_{\a_{1}\a_{2}}^{\b_{1}\b_{2}} \cdots R_{\a_{2n-1}\a_{2n}}^{\b_{2n-1}\b_{2n}}.
\ee
Then, if we consider the coupling, 
\be
{\cal L}_{2 n H}=\sqrt{-g}H_{\m}^{\,\,\,\n}\nabla^{\m}\phi\nabla_{\n}\phi,
\ee
varying with respect to the metric gives us 
\bea
&&\frac{\delta {\cal L}_{2n H}}{\sqrt{-g}\,\delta g^{\m\n}}=-\frac{1}{2}g_{\m\n}H_{\a} \,^{\b} \nabla^{\a}\phi\nabla_{\b}\phi+H_{(\m} \,^{\r} \nabla_{\n)}\phi\nabla_{\r}\phi\nonumber \\
&&\qquad \qquad \qquad-\frac{n}{2^{n+1}}\d^{\lambda \a_{1}\a_{2} \cdots \a_{2n-1}\a_{2n}}_{\k \b_{1}\b_{2} \cdots \b_{2n-1} (\m}R^{\b_{2n-1}}_{\,\,\,\,\,\,\,\,\,\,\,\,\,\,\n)\a_{2n-1}\a_{2n}}R^{\b_{1}\b_{2}}_{\a_{1}\a_{2}} \cdots R^{\b_{2n-3}\b_{2n-2}}_{\a_{2n-3}\a_{2n-2}}\nabla^{\k}\phi\nabla_{\lambda}\phi\nonumber \\
&&\qquad \qquad \qquad-\frac{n}{2^{n+1}}\delta^{\lambda \a_{1}\a_{2} \cdots \a_{2n-3}\a_{2n-2}\a_{2n-1}\a_{2n}}_{\k \b_{1}\b_{2} \cdots \b_{2n-3}\b_{2n-2}\,\,\,\,\,\,\b_{2n}\,\,\,\,\,\, (\m}g_{\n)\a_{2n}}R^{\b_{1}\b_{2}}_{\a_{1}\a_{2}} \cdots R^{\b_{2n-3}\b_{2n-2}}_{\a_{2n-3}\a_{2n-2}} R_{\a_{2n-1}}\,^{\r\b_{2n}\k}\nabla_{\r}\phi\nabla_{\lambda}\phi\nonumber \\
&&\qquad \qquad \qquad-\frac{n}{2^{n}}\delta^{\lambda \a_{1}\a_{2} \cdots \a_{2n-3}\a_{2n-2}\a_{2n-1}\a_{2n}}_{\k \b_{1}\b_{2} \cdots \b_{2n-3}\b_{2n-2}\,\,\,\,\,\,\b_{2n} \,\,\,\,\,\,(\m}g_{\n)\a_{2n}}R^{\b_{1}\b_{2}}_{\a_{1}\a_{2}} \cdots R^{\b_{2n-3}\b_{2n-2}}_{\a_{2n-3}\a_{2n-2}} \nabla^{\k}\nabla_{\a_{2n-1}}\phi\nabla^{\b_{2n}}\nabla_{\lambda}\phi.
\eea

\end{appendix}

%%%%%%%%%%%%%%%%%%%%%%%%%%%% BIBLIOGRAPHY %%%%%%%%%%%%%%%%%%%%%%%%%%%%%%%%%%%%%%

\end{document}